\documentclass[12pt]{article}
\usepackage{amssymb}
\usepackage{epsfig}
\usepackage[
            body={14.6true cm,22true cm},
            ]{geometry}
\begin{document}

\title{\bf The macro-behavior of agents' opinion under the influence of an external field}

\author{{YunFeng Chang\footnote{Email: yunfeng.chang@gmail.com}, Long Guo, Xu Cai}\\
{\small Complexity Science Center,Institute of Particle Physics}\\
{\small Huazhong(Central China) Normal University, Wuhan, 430079,
China.}} \maketitle

\begin{abstract}
In this paper, a model about the evolution of opinion on small
world networks is proposed. We studied the macro-behavior of the
agents' opinion and the relative change rate as time elapses. The
external field was found to play an important role in making the
opinion $s(t)$ balance or increase, and without the influence of
the external field, the relative change rate $\gamma(t)$ shows a
nonlinear increasing behavior as time runs. What's more, this
nonlinear increasing behavior is independent of the initial
condition, the strength of the external field and the time that we
cancel the external field. Maybe the results can reflect some
phenomenon in our society, such as the function of the
macro-control in China or the Mass Media in our society.
\end{abstract}

\bigskip

\section{Introduction}

Since Watts and Strogatz's work\cite{s1} on small-world network in
1998 and Barab$\acute{a}$si and Albert's work\cite{s2} on
scale-free network in 1999, an explosion of work about complex
networks emerges, from the analysis of the topology of real
networks\cite{s3,s4,s5} to the evolution dynamics of complex
networks, and using complex networks to model all kinds of complex
systems\cite{s1,s2} and different kinds of dynamical
processes\cite{s6,s7,s8}, and with special interest on how does
the network structure affect the properties of a dynamical system.
In this paper we will focus on the dynamical process of the
evolution of opinion on small world social networks under the
influence of an external field.

Many models about opinion dynamics have been proposed in recent
years. At first, only binary opinion models were
considered\cite{s9,s10,s11,s12}, some of which used "social impact
theory"\cite{s13,s14,s15} founded by Latan$\acute{e}$ to describe
the transition from private attitude to public opinion, which can
help us understand minority or majority consensus and the
formation of the phenomenon that many agents sharing the same
opinion\cite{s10}. Then people extended the models to continuous
opinion models\cite{s13,s16,s17}. However, they are all "Bounded
Confidence" models, which means they all set a parameter called
"bounded confidence" as a threshold for the updating of opinion,
i.e., two agents will interact with each other only when their
opinions are close enough. Take Deffuant et al.(D)
model\cite{s13,s17,s18} and Hegselmann and Krause(HK)
model\cite{s19} for example, they study how the damage spreading
and the fraction of perturbed agents vary with "bounded
confidence" and time.

In this paper, we proposed a new model with the "bounded
confidence" omitted to investigate the opinion evolution on small
world networks under the influence of an external field. The
external field here plays some role like the macro-control policy
of Chinese government or the Mass Media for example, and we focus
on the macro-behavior of agents' opinion evolution under the
influence of such an external field. The results show that the
macro-behavior of the agents' opinion get balanced after some time
steps under the interaction between the topology of the network
and the external field, and the time when the system reaches the
maximum opinion has a power law relationship with the power of the
external field. What's more, if we cancel the external field at
one certain step, the relative change rate of the macro-behavior
of the agents' opinion shows a nonlinear increasing behavior, and
this behavior is independent of the initial condition, the
strength of the external field and the time that we cancel the
external field.

In the next section we will propose our model. In section 3, we
will give out the results and finally in section 4, our
conclusions.

\section{The Model}

A social network is a set of agents or groups with relationships
of different kinds among them\cite{s20,s21}, such as friendship,
collaboration, business, sexual and other interactions. Thinking
about the question how do you get or change your information about
things that you care? Most probably, you exchange your opinion
with your friends and get information from the TV, newspaper, even
the government policy and so on. Our model is dedicate to describe
this opinion evolution process and is defined in the following
way: the social relationship among agents is mapped onto a complex
network, the nodes in the network represent the agents in the
society and the edges represent the relationship because of which
the agents can exchange their information to update their opinions
on something.

When simulating we base our model on a network with $N$ nodes (the
agents) and K edges. We represent the network by an $N\times N$
adjacency matrix $a_{ij}$. The element $a_{ij}$ of this matrix is
$0$ if agent $i$ and agent $j$ can not be influenced by each
other, and $a_{ij}$ is $1$ if agent $i$ and agent $j$ can be
influenced by each other. The strength of agent $i$'s opinion
$s_{i}(t)$ at time $t$  (can be thought as the degree of one's
concerning about one certain thing) is a real number in the range
of [0,1), which varies under the influence of its nearest
neighbors and the external field, that means the agents update
their opinions according to the influence of their nearest
neighbors and the external field. We will give out the definition
of the quantities in our model in the following.

First, the quantity:
\begin{equation}
{\rm
Inf}^{(i)}_{j}(t)=\frac{k_{j}}{\sum\limits_{<i,l>}^{}k_{l}}s_{j}(t)
\end{equation}
$\sum\limits_{<i,l>}^{}$ sums over all the nearest neighbors of
agent $i$. ${\rm Inf}^{(i)}_{j}(t)$ represents the influence of
agent $j$ on agent $i$ at time $t$, which is related to $j$'s
opinion $s_{j}(t)$, $j$'s degree $k_{j}$ and the total degree of
the nearest neighbors of agent $i$.

Second, every agent has a different role in the network, which can
be described as something like the "charge" of agent $i$:
\begin{equation}
e_{i}=\frac{k_{i}}{\sum\limits_{j}^{}k_{j}}
\end{equation}
$\sum\limits_{j}^{}$  sums over all the agents.

Third, we define the average opinion of agent $i$'s neighbors at
time $t$:
\begin{equation}
\overline{s_{i}(t)}=\frac{\sum\limits_{<i,j>}^{}k_{j}s_{j}(t)}{\sum\limits_{<i,l>}^{}k_{l}}
\end{equation}
$\sum\limits_{<i,j>}^{}$ and $\sum\limits_{<i,l>}^{}$ sum over all
the nearest neighbors of agent $i$. $\overline{s_{i}(t)}$ can
describe the opinions of the local community around individual
$i$.

Finally, we define the "power" $Q$ of the external field which
reflects the strength of the influence of the external field, $Q$
acts on all the agents. Then the influence of the external field
to individual $i$ is:
\begin{equation}
Qe_{i}=Q\frac{k_{i}}{\sum\limits_{j}^{}k_{j}}
\end{equation}

There are many real factors that influence agents to change their
opinions about things in our society, such as mass media or
government polices. We consider these cases as the influence of an
external field. Under the influence of the external field, agents
update their opinions according to the following equations, when
$s_{i}(t)=0$,
\begin{equation}
s_{i}(t+1)=\left\{\begin{array}{clcc}
\frac{1}{k_{i}}\sum\limits_{<i,j>}^{}{\rm
Inf}^{(i)}_{j}(t)+Qe_{i}&,&{with
prob 50\%}\\
0&,&{with prob 50\%}
\end{array}
\right.
\end{equation}
and when $s_{i}(t)\neq0$,
\begin{equation}
\begin{array}{lll}
\displaystyle s_{i}(t+1)&=& s_{i}(t)+\Theta(\overline{s_{i}(t)}\\
& &-s_{i}(t))\frac{s_{i}(t)}{k_{i}}\sum\limits_{<i,j>}^{}{\rm Inf}^{(i)}_{j}(t)\\
& &+\Theta(Q-s_{i}(t))Qe_{i}s_{i}(t)
\end{array}
\end{equation}
where
\begin{equation}
\Theta(x)=\left\{\begin{array}{clcc}
1&,\hspace{0.4cm}& {x\geq0}\\
-1&,&{x<0}
\end{array}
\right.
\end{equation}
is the step function. The step function means the neighbors of
agent $i$ (or the external field) can have negative influence on
$s_{i}(t)$ if $\overline{s_{i}(t)}$ (or $Q$) is weaker than
$s_{i}(t)$ as well as positive influence while
$\overline{s_{i}(t)}$ (or $Q$) is stronger than $s_{i}(t)$ at time
$t$. And the coupling of $s_{i}(t)$ and $s_{j}(t)$ (or $Q$)
indicates $i$'s intention to keep its opinion unchanged. This
interaction process is similar to the interaction among people in
real life: individual often changes his opinion about things due
to the influence of his surrounding friends and some other
external factors but intends to keep his opinion unchanged at the
same time.

\section{Results and Analysis}
We realize our model on small world networks with scale free
degree distribution since small world and scale free property are
the most common properties of social networks. The network used in
this paper has $N=400$ agents and its average degree $<k>=29.755$,
the characteristic path length $L=2.13$ and the clustering
coefficient $C=0.241$. What's more, when initialize the network
agents with their opinions, we control the initial process to let
the initial average opinion to be a small value since we want to
know how the topology of the network and the external field
interact with each other and how they affect the opinion
evolution.

Fig.1 is the results of the evolution of the macro-behavior
$s(t)=\frac{1}{N}\sum\limits_{i}^{}s_{i}(t)$ under different
external field (Fig.1a) and with different initial condition
(Fig.1b) as time elapses. We can see that they all have the same
evolution trend. We can divide the evolution process into three
sections: at first, the interaction among the agents controls the
evolution process (about the first 100 time steps), in this
process the opinion changes slow; after a short time steps the
macro-behavior of the agents' opinion increases faster than ever
(about $100^{th}-500^{th}$ time steps), in this process the
external field plays an more important role than the topology of
the network, and arrived at the extreme value. Thirdly, the
macro-behavior of the agents' opinion get balanced under the
interaction between the topology of the network and the external
field, in this section the influence of the topology of the
network has the same importance as that of the external field. And
we can see that the opinion changes quickly under the influence of
the external field, which reflects the strong role of the external
field on the social opinion evolution. What's more, the time when
the system reaches the extreme value has power law relationship
with the power of the external field (Fig.1c). This may reflect
that you can accelerate the process of getting the best impact by
improving your investment on the power of your policy or media.

\begin{figure}
\begin{center}
\includegraphics[width=1.0\textwidth]{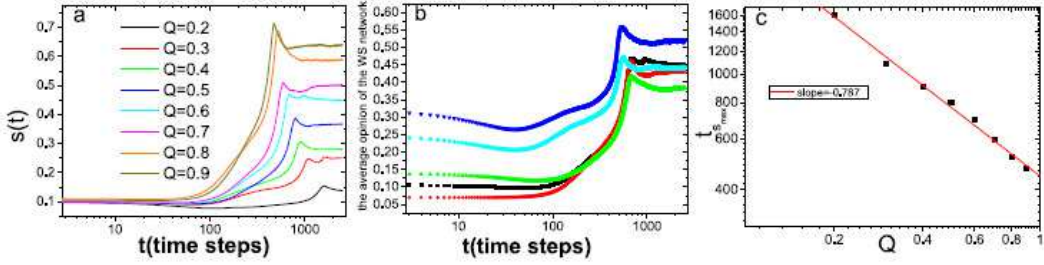}
\caption{the macro-behavior of the agents' opinion under
   different external field (a) and with different initial
   condition (b), the relationship between the time when the opinion get extreme value and the power of the external field (c), (a) has the same initial
   opinion $s(0)=0.10\pm0.06$ , and (b) has the same external field
   $Q=0.6$}
\end{center}
\end{figure}

Then we cancel the influence of the external field at one certain
time step during the simulation. We find that, without the
influence of the external field, the evolution of the
macro-behavior of the agents' opinion shows an exponentially decay
behavior (see Fig.2). That is to say, with only the interaction
among the nearest neighbors, the agents' opinion will fade away
exponentially as time elapses.

\begin{figure}
\begin{center}
\includegraphics[width=0.6\textwidth]{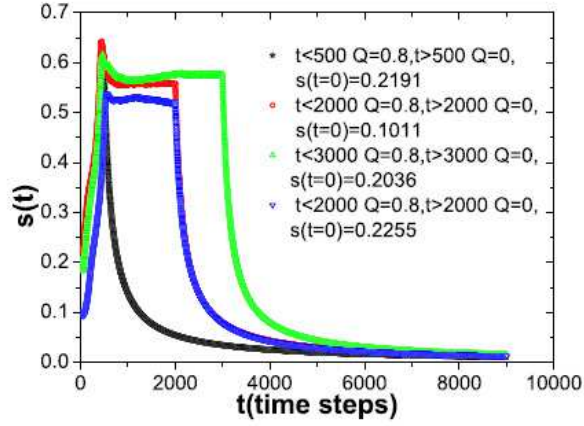}
\caption{the evolution of the macro-behavior of
   the agents' opinion as time elapses under the different initial
   condition and the same external field, but cancel the influence of the external time at different time.}
\end{center}
\end{figure}

\begin{figure}
\begin{center}
\includegraphics[width=0.8\textwidth]{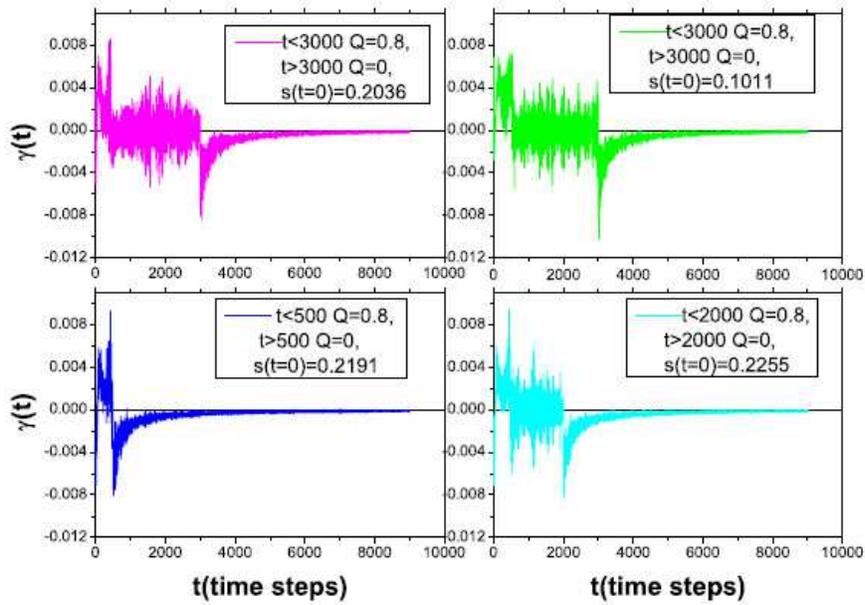}
\caption{the relative change rate of the
   macro-behavior of the agents' opinion varies as time elapses
   with different initial condition, the same
   external field and at different time that we cancel the
   influence of the external field.}
\end{center}
\end{figure}

Thirdly, we defined the quantity: the relative change rate
$\gamma(t)=\frac{s(t+1)-s(t)}{s(t)}$,
$s(t)=\frac{1}{N}\sum\limits_{i}^{}s_{i}(t)$ is the average
opinion of the whole network, i.e. the macro-behavior of the
agents' opinion at time $t$. $\gamma(t)$ is used to measure the
relative change of the macro-behavior of the agents' opinion at
time t. We do computer experiment with different initial condition
, under different external field and at different time step that
we cancel the influence of the external field (see Fig.3). We get
the result that the influence of the agents' interaction on the
opinion's evolution is unrelated to the initial condition, the
strength of the external field and the time when we cancel the
influence of the external field, it is only related to the
topology of the network. After the cancel of the external field,
the relative change rate shows nonlinear increasing as time
elapses (see Fig.3), but below the line $\gamma(t)=0$. On the
other hand, seeing the first part of the four graphs in Fig.3, we
can see that the competition between the interaction among agents
and the external field makes the macro-behavior of the agents'
opinion balance. This reflects the results in Fig.1 and the
important role of the external field on the network's opinion
evolution, which can make the opinion balance or increase.

\section{Conclusion}

In this paper we proposed a model to describe the phenomenon of
the continue opinion evolution on small-world networks. We studied
the macro-behavior $s(t)$ of the agents' opinion, the relationship
between the time when system reaches the maximum opinion and the
power of the external field, and the relative change rate
$\gamma(t)$ as time elapses. We find that the external field has
more important role in making the opinion balance or increase (see
the pictures above mentioned) and the time the system reaches the
maximum opinion has power law relationship with the power of the
external field. After we cancel the external field, the relative
change rate shows nonlinear increasing as the time runs, which is
related only to the topology of the network. Maybe the results can
reflect some phenomenon in our society, such as the function of
the macro-control in China or the Mass Media in our society for
example. Compared with the most recent work\cite{s22} on the study
of opinion evolution, [22] focus on the coevolution of multi-kinds
of different opinions to see the opinion formation and the
formation of communities, while our model focus on the
macro-behavior of agents' opinion under the influence of the
external field. On the other hand, from the result of our model,
we can see that the evolution of opinion on networks has a strong
relationship with the topology of the networks, we will go on our
study about the evolution of opinion on different kinds of
networks in our future work.

\section*{Acknowledgements}

This work is supported by the National Natural Science Foundation
of China under Grant NOs 70571027, 10647125 and 70401020.

\end{document}